\documentclass[onecolumn,showpacs,preprintnumbers,elsart]{revtex4}
\usepackage{amsmath}
\usepackage{mathrsfs}
\usepackage{graphicx}
\usepackage{dcolumn}
\usepackage{bm}
\usepackage[center]{subfigure}
\usepackage{color}

\begin{document}

\title{Switch between the types of the symmetry breaking bifurcation in optically induced photorefractive rotational double-well potential}
\author{Guihua Chen$^{1}$}
\email{cghphys@gmail.com; Tel/fax numbers: +86-0769-22861268}
\author{Jianxiong Wu$^{2}$}
\author{Zhihuan Luo$^{3}$}

\affiliation{$^{1}$Department of electronic engineering, Dongguan University of Technology, Dongguan, 523808, China\\
$^{2}$Department of electrical and Engineering, University of Toronto, Toronto, M5S3G4, Canada\\
$^{3}$State Key Laboratory of Optoelectronic Materials and Technologies,\\
Sun Yat-sen University, Guangzhou 510275, China}

\begin{abstract}
 We study the possibility of switching the types of symmetry breaking bifurcation (SBB) in the cylinder shell waveguide with helical double-well potential along propagation direction. This model is described by the one-dimensional nonlinear Schr\"{o}dinger (NLS) equation. The symmetry- and antisymmetry-breakings can be caused by increasing the applied voltage onto the waveguide in the self-focusing and -defocusing cases, respectively. In the self-focusing case, the type of SBB can be switched from supercritical to subcritical. While in the self-defocusing case, the type of SBB can not be switched because only one type of SBB is found.
\end{abstract}

\maketitle



Phase transition is a fundamental topic in many branches of physics. Spontaneous symmetric breaking (SSB), which is the transition from the symmetric ground state (GS) to that which does not follow the symmetry of the potential, always plays an important role in inducing the transition \cite{Landau}. In nonlinear systems, because additional nonlinearity always give rise to the effect of SSB, phase transition and SSB are also important issues and attract great attention. Among many kinds of nonlinear systems, double-well potential (DWP) or dual-core system is the most essential model employed to study the phase transition and SSB of the nonlinear states \cite{Kevredidis,Theocharis,Albuch,Herring,Matuszewski,Trippenbach,Satija,Xiong}. In DWP system, the important process relating to the phase transition and SSB is symmetric breaking bifurcation (SBB), which determines the process of symmetric states transiting to the asymmetric ones \cite{Mayteevarunyoo,Salasnich,Adhikari,Bar,Sakaguchi,Brazhnyi,Davoyan,YLi}. There are two kinds of SBB, subcritical and super critical, which are tantamount to the phase transition of the first and second kind respectively. The former kind of SBB, subcritical type, features branches of the asymmetric states going at first backward after the bifurcation point then turning forward. While the latter one, the supercritical type, features the asymmetric branches going immediately to the forward direction after the bifurcation point. Different settings with different environments may possibly lead to different types of SBB. An interesting problem is the possibility to control the type of the SBB, and thus to switch between the respective phase transitions of the two kinds. About this problem, it has been reported that the addition of a periodic potential (optical lattice) acting in the unconfined direction changes the character of the SBB from sub- to supercritical \cite{Trippenbach}. Recently, it is also reported that the periodical modulation of the linear-coupling constant, which is induced by the electromagnetically induced transparency (EIT) via the double-$\Lambda$ system, can change the SBB from sub- to supercritical with the increase of the total power of the probe beams \cite{YLi}.

Very recently, a setting with rotational (alias helical) DWP potential in a cylinder shell waveguide was performed \cite{YLi2}, which reported that the rotating speed of the potential could also induce the symmetry breaking of the nonlinear modes. It is well known that rotating systems often provide more interesting dynamics than their stationary counterparts \cite{Kakarantzas,Saito,Schwartz,Kartashov,SJia,Fetter,LWen,Sakaguchi2}. A natural problem is, whether we can control the type of SBB by means of the rotating speed of the potential or not?
\begin{figure}[tbp]
\centering%
{\label{fig1a}
\includegraphics[scale=0.4]{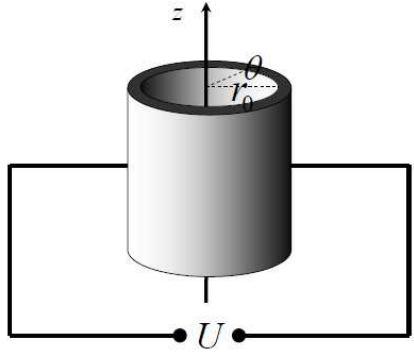}}%
\caption{Biased photorefractive cylinder waveguide shell, $r_{0}$ is the radius of the circular and $U$ is the applied biased voltage onto the crystal.}
\label{Fig1}
\end{figure}

In this work, we build a cylinder waveguide shell with a helical DWP by means of the biased photorefractive crystals (BPRCs) (See in Fig. \ref{Fig1}) to study the controllability of type of SBB in the rotating system. The BPRC, which features a saturable nonlinearity, is a kind of important nonlinear optical material, especially in creating the spatial optical solitons. In the past decades, many kinds of spatial solitons are observed from BPRCs \cite{Efremidis,Fleischer,Buljan,Wang,Lederer,PZhang,YLi3,ZChen}. As illustrated in the Ref. \cite{YLi2}, helical DWP in the waveguide shell can be created by means of optical-induction techniques. For example, using two waves, which have the same amplitude but opposite vortical pumps, ordinary polarizations in a photorefractive material, and a very small mismatch of the propagation constant, can form a helical DWP. If the width of the shell is thin enough, the problem of this type can be reduced to a quasi one-dimensional (1D) with the argument chosen to be the azimuth angle $\theta$, which require the equation obeying the periodical boundary condition. The propagation dynamics of the probe wave, which is launched in the extraordinary polarization of the BPRC, can be described by the underlying 1D scaled nonlinear Schr\"{o}dinger equation in the rotating reference frame,
\begin{eqnarray}
i{\partial\over\partial z}\psi=-{1\over2r^{2}_{0}}{\partial^{2}\over\partial\theta^{2}}\psi+i\omega{\partial\over\partial\theta}\psi+{E_{0}\over1+I+|\psi|^{2}}\psi,\label{NLS}
\end{eqnarray}
where $r_{0}$, which can be fixed to $1$, is the radius of the cylinder shell, $\omega$ is the rotating speed of the potential, $I$ is the induction field (alias the pumps) and $E_{0}$ is the nonlinear parameter which is defined by the magnitude and polarity of the applied biased voltage $U$. The periodical boundary condition requires the probe and the pump satisfying $\psi(\theta)\equiv \psi(\theta\pm2k\pi)$ and $I(\theta)\equiv I(\theta\pm2k\pi)$ $(k=1,2,3\ldots)$. The total power of the probe $P$ and its stationary mode with real propagation constant $\beta$,
\begin{eqnarray}
P=\int^{\pi}_{-\pi}|\psi|^{2}d\theta, \label{Power}
\end{eqnarray}
\begin{eqnarray}
\psi(\theta,z)=e^{i\beta z}\phi(\theta). \label{solution}
\end{eqnarray}
From the Ref. \cite{YLi2}, if the rotating speed $\omega\neq0$, $\phi(\theta)$ is a complex function. The stability of the stationary mode can be numerical identified by the computation of eigenvalues for small perturbations and direct simulations. The perturbed solution is given as $\psi=e^{i\beta z}[\phi(\theta)+u(\theta)e^{i\lambda z}+v^{\ast}(\theta)e^{-i\lambda z}]$. Substitution of this ansatz into Eq. (\ref{NLS}) and linearization lead to the eigenvalue problem,
\begin{equation}
\left[
\begin{array}{cc}
{1\over2}{d^{2}\over d\theta^{2}}-i\omega{d\over d\theta}-\beta-{E_{0}\over F}\left(1-{|\phi|^{2}\over F}\right) &
{E_{0}\over F}\phi^{2} \\
{E_{0}\over F}\phi^{\ast 2} & -{1\over2}{d^{2}\over d\theta^{2}}-i\omega{d\over d\theta}+\beta+{E_{0}\over F}\left(1-{|\phi|^{2}\over F}\right)
\end{array}%
\right] \left(
\begin{array}{c}
u \\
v%
\end{array}%
\right) =\lambda \left(
\begin{array}{c}
u \\
v%
\end{array}%
\right) ,  \label{lambda}
\end{equation}%
where $F=1+I+|\phi|^{2}$. The solution $\phi$ is stable if all the eigenvalues of Eq. (\ref{lambda}) are real.

\begin{figure}[tbp]
\centering%
\subfigure[] {\label{fig_2_a}
\includegraphics[scale=0.21]{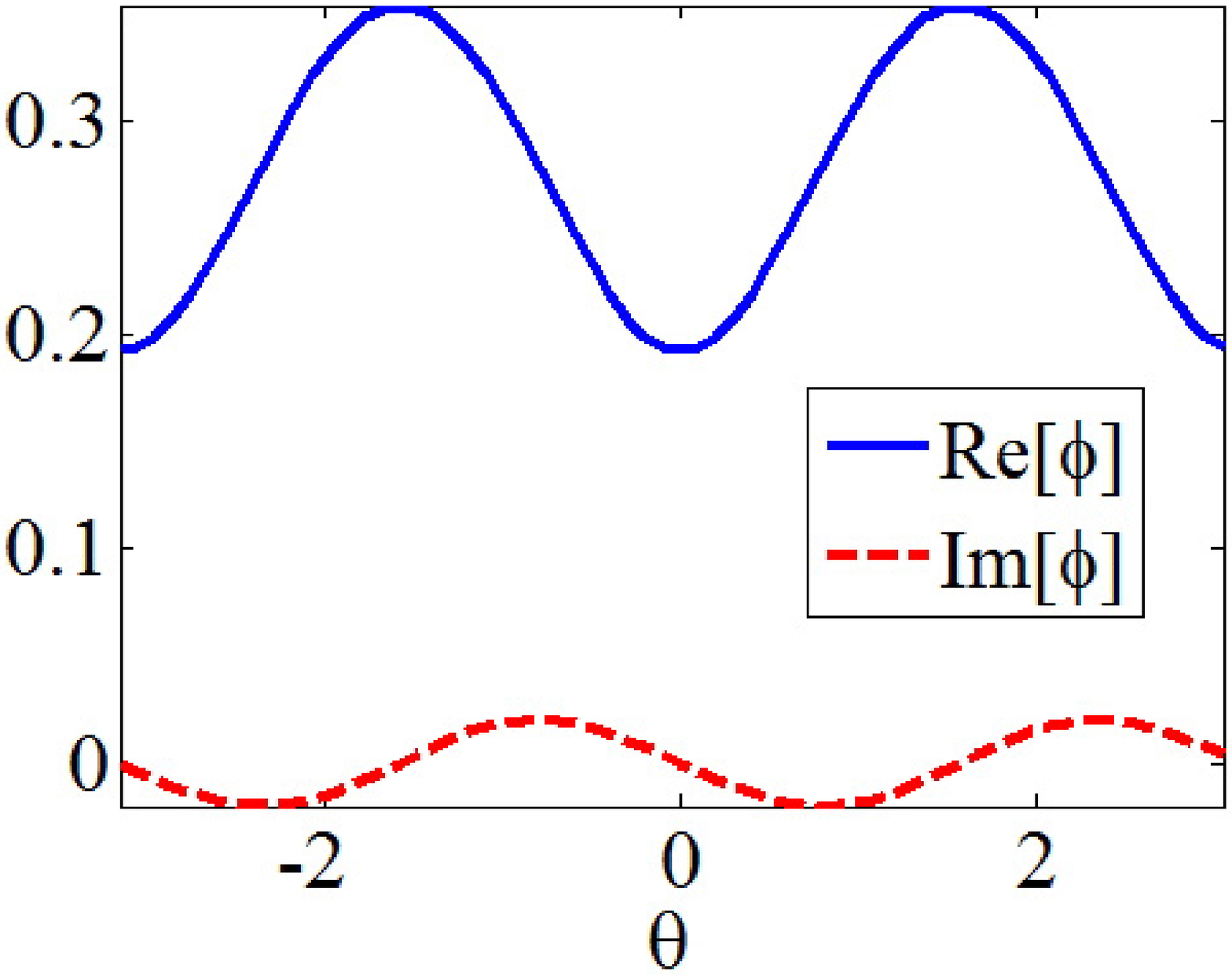}}
\subfigure[] {\label{fig_2_b}
\includegraphics[scale=0.21]{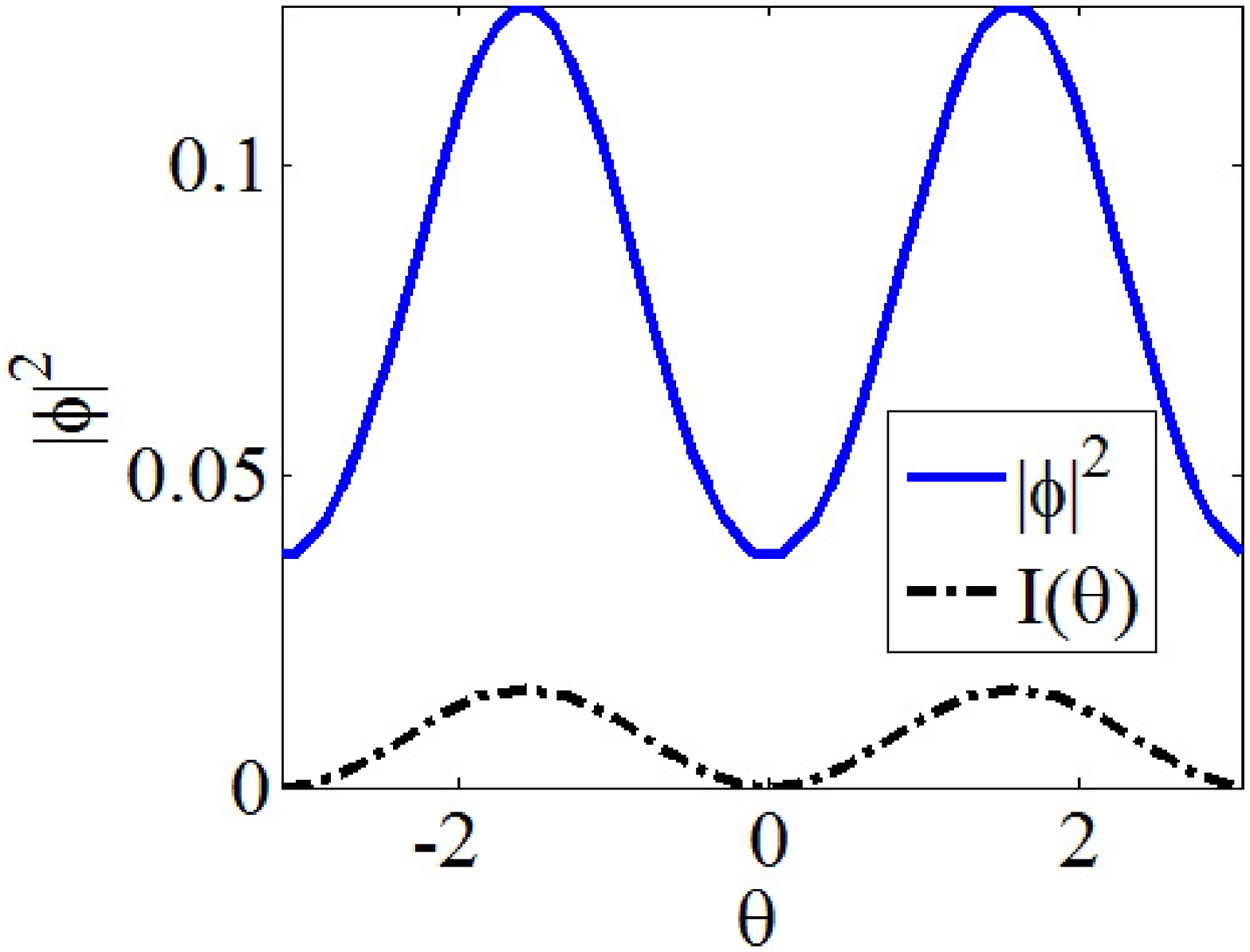}}
\subfigure[] {\label{fig_2_c}
\includegraphics[scale=0.21]{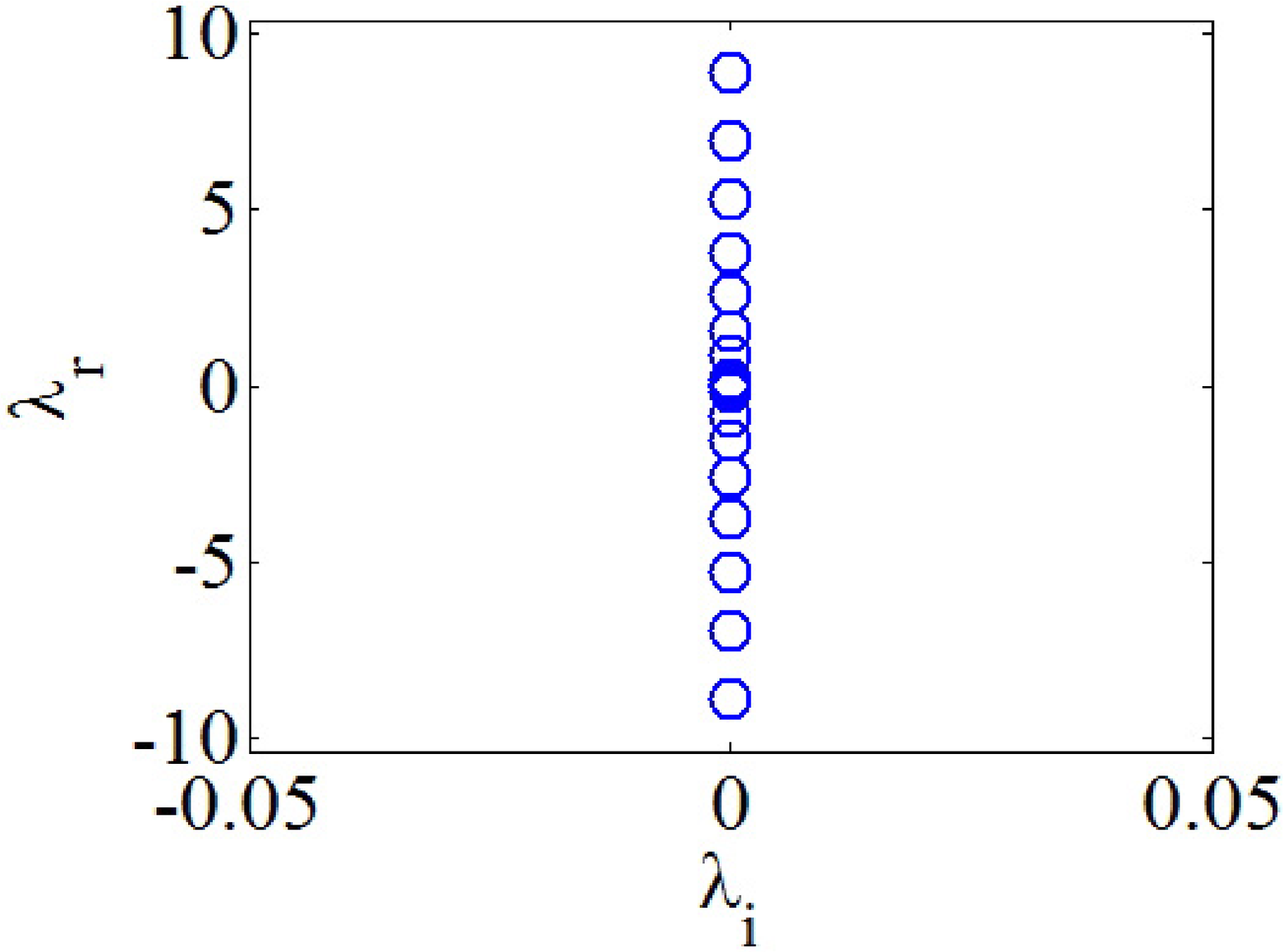}}
\subfigure[] {\label{fig_2_d}
\includegraphics[scale=0.21]{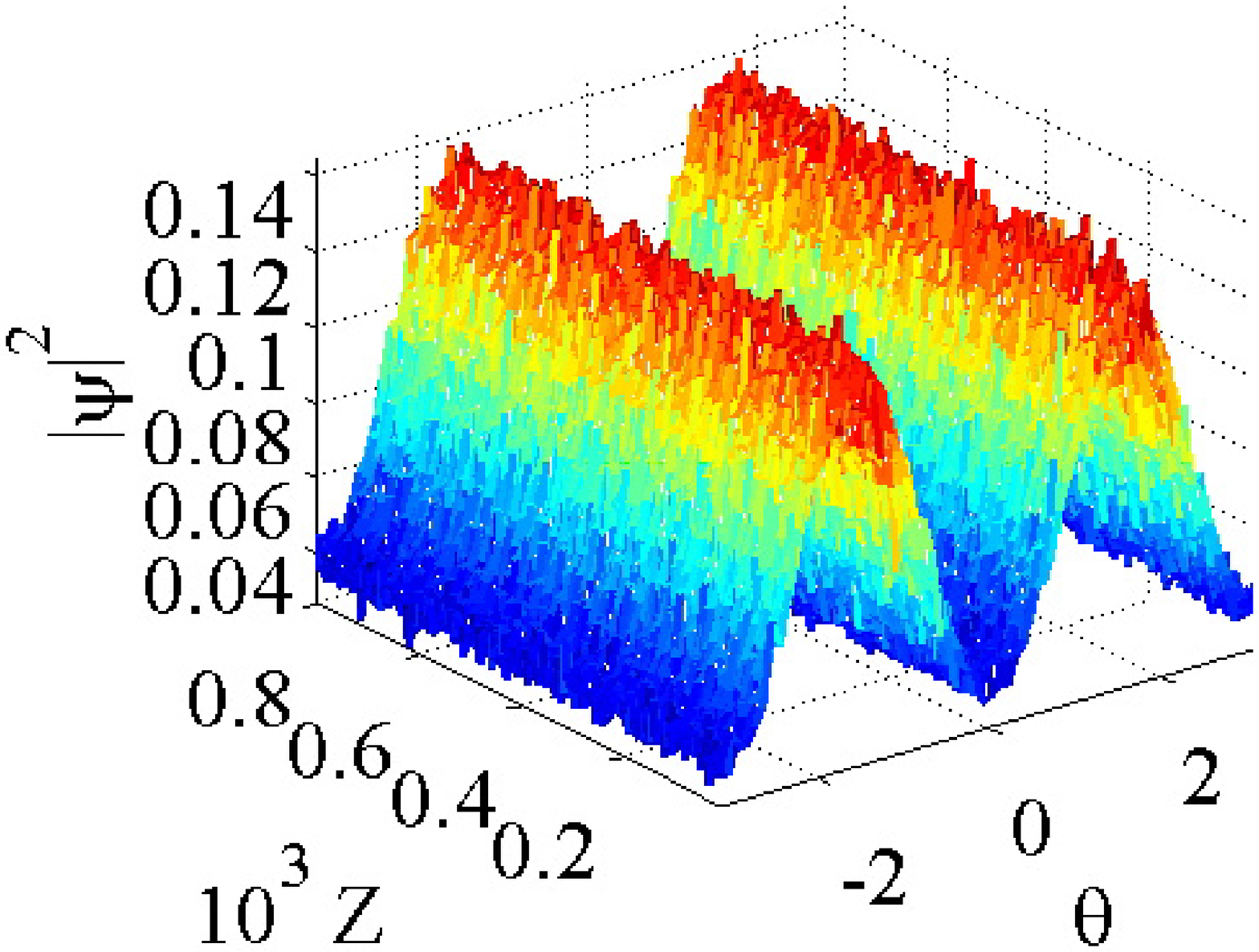}}
\subfigure[]{ \label{fig_2_e}
\includegraphics[scale=0.21]{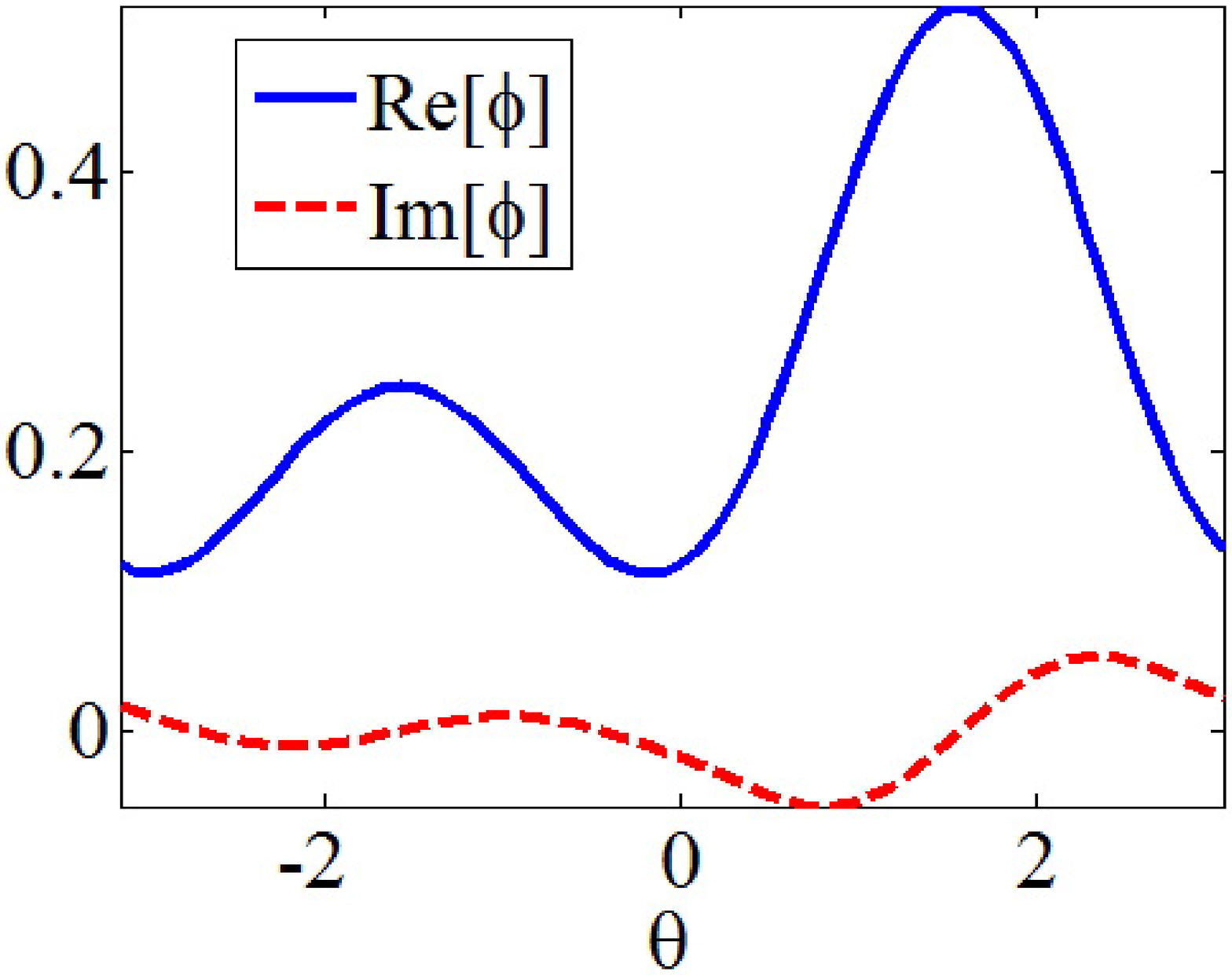}}
\subfigure[]{ \label{fig_2_f}
\includegraphics[scale=0.21]{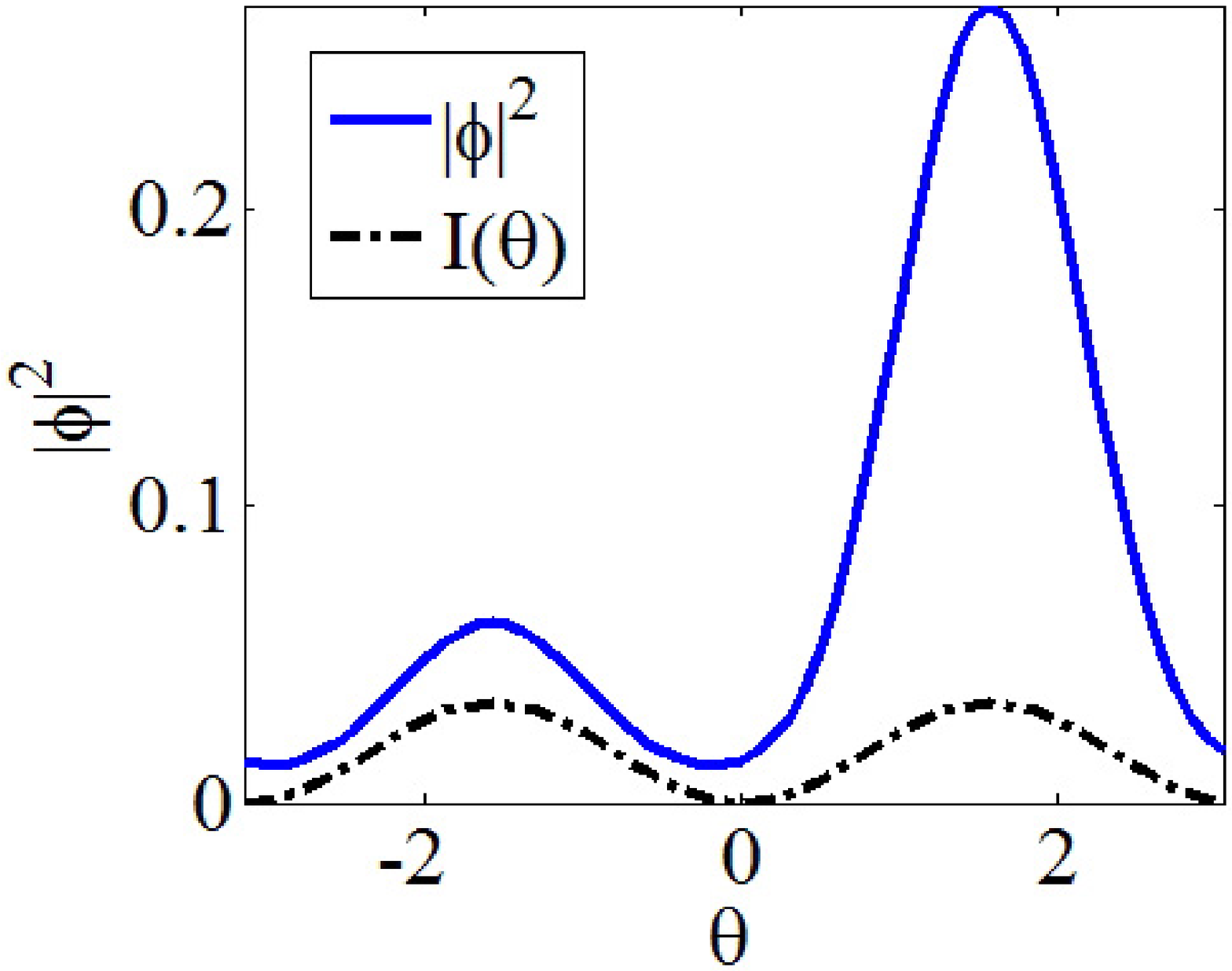}}
\subfigure[] {\label{fig_2_g}
\includegraphics[scale=0.21]{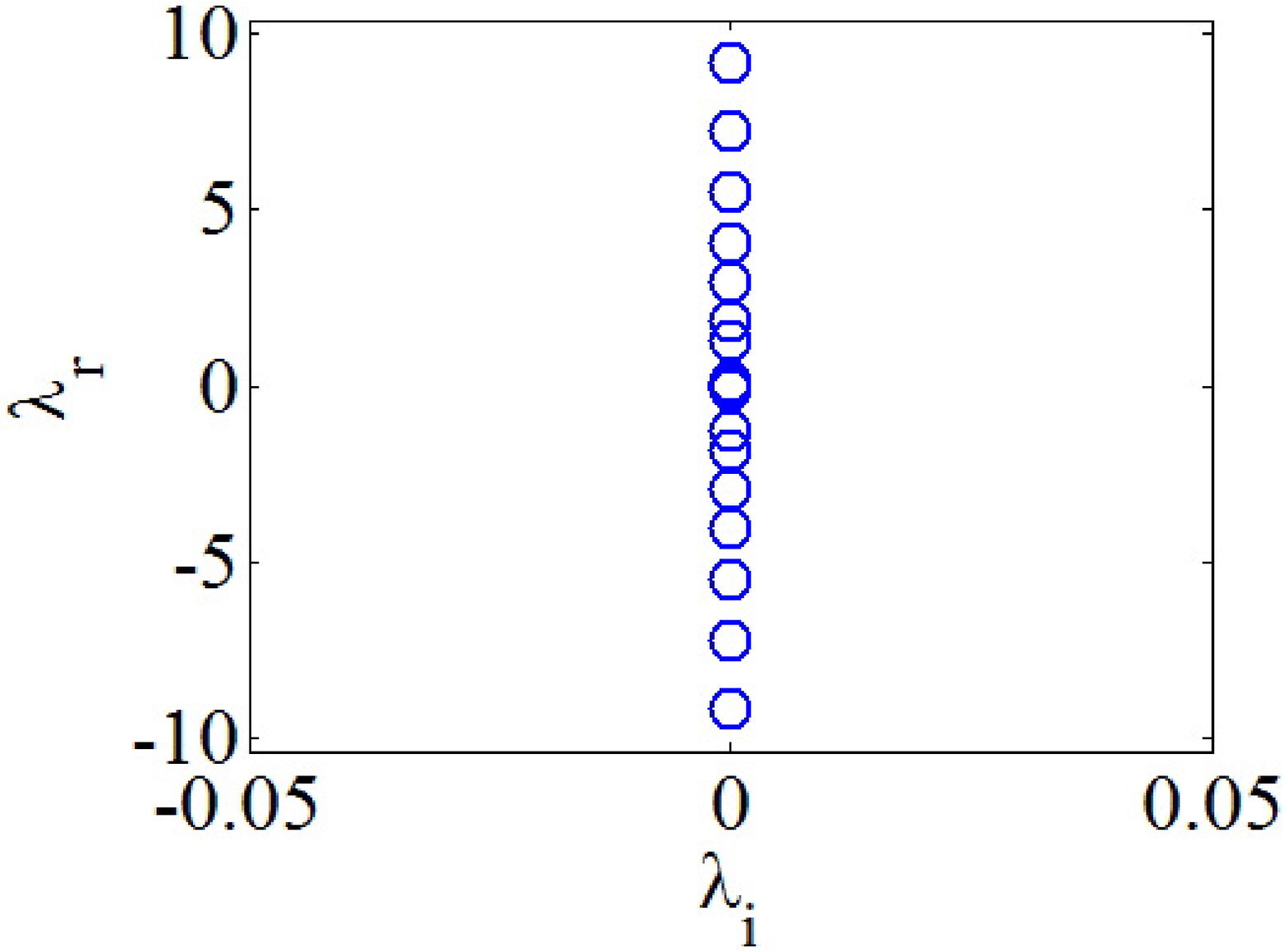}}
\subfigure[] {\label{fig_2_h}
\includegraphics[scale=0.21]{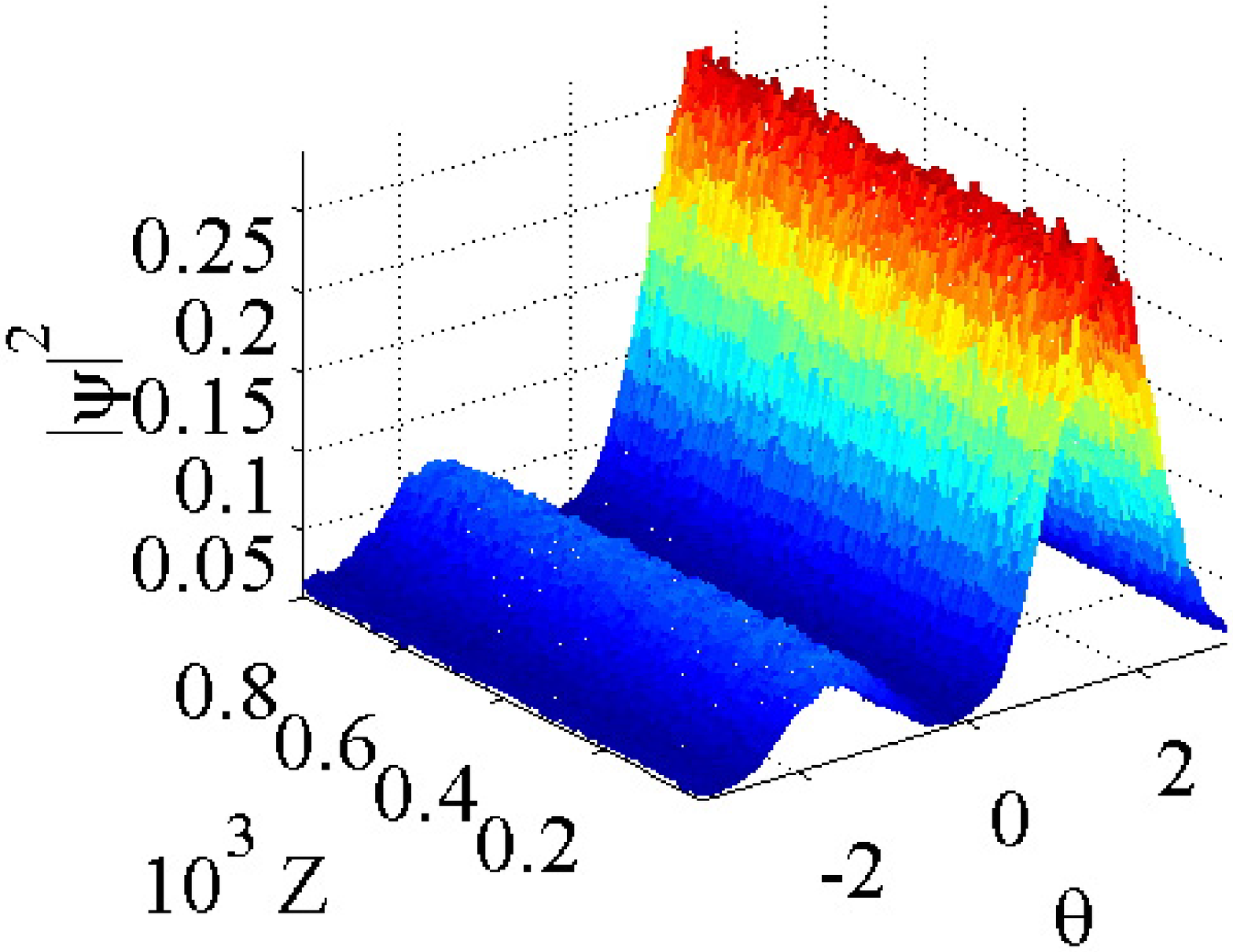}}
\caption{(Color online) Examples of stable symmetric and asymmetric GS modes,
found in the system with SF case at rotation speed $\omega =0.25$, with the parameters $(E_{0},A,P)=(2,2,0.5)$ and $(4,2,0.5)$. Panels (a), (e) display, severally, real and imaginary parts of the symmetric and asymmetric GS modes, while (b), (f) show their local-power (density) profiles. The dashed-dot curves in (b) and (f) depict the corresponding intensity profile, $I(\theta)$; in the present case, it is $\sin ^{2}\protect\theta$. Panels (c), (g) exhibit the growing rates of the wave modes. And panel (d), (h) demonstrate the evolutions of these wave modes perturbed by 10\% noises, respectively. }
\label{GSSF}
\end{figure}

\begin{figure}[tbp]
\centering%
\subfigure[] {\label{fig3a}
\includegraphics[scale=0.2]{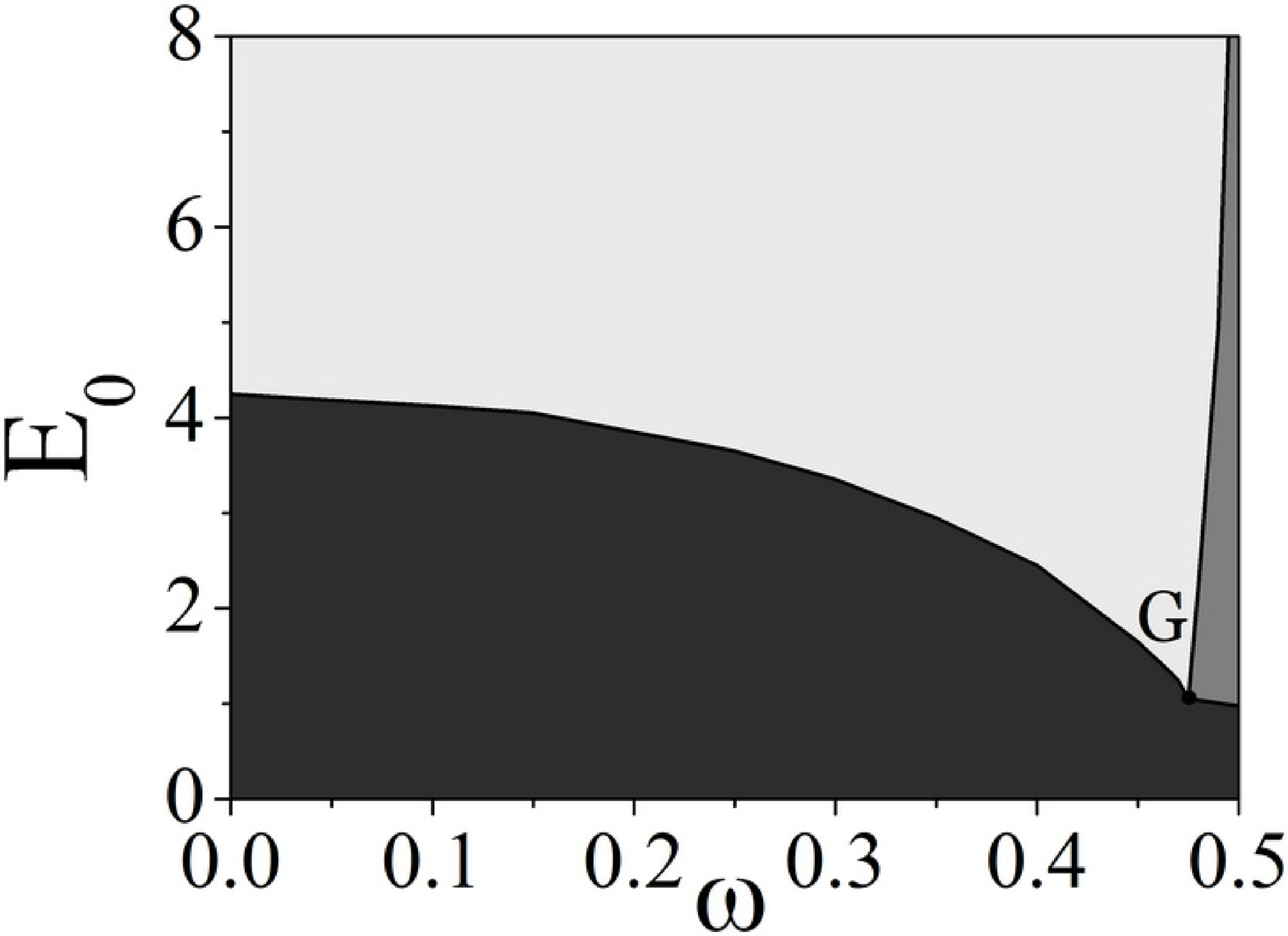}}%
\subfigure[] {\label{fig3b}
\includegraphics[scale=0.2]{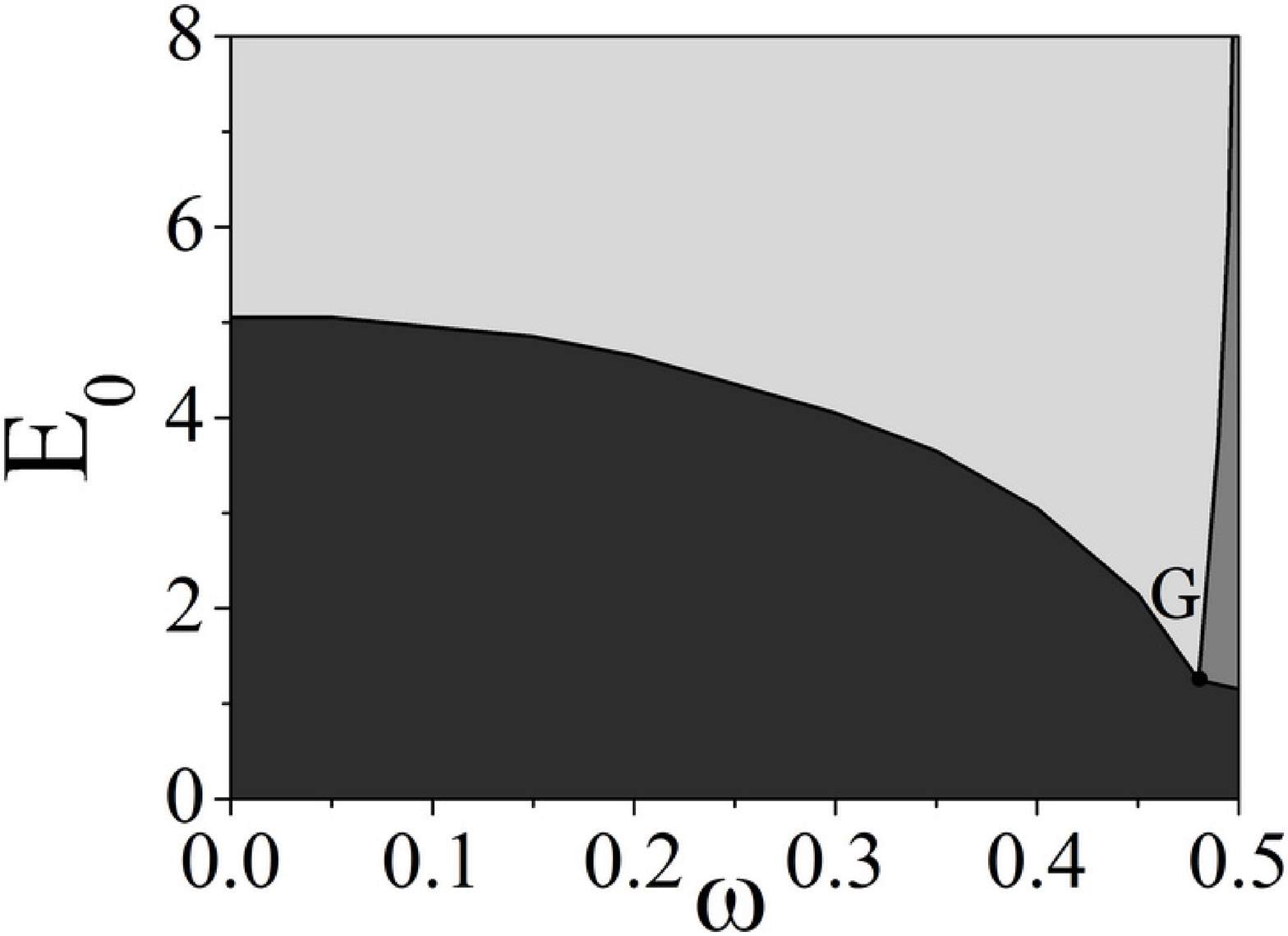}}
\subfigure[]{ \label{fig3c}
\includegraphics[scale=0.2]{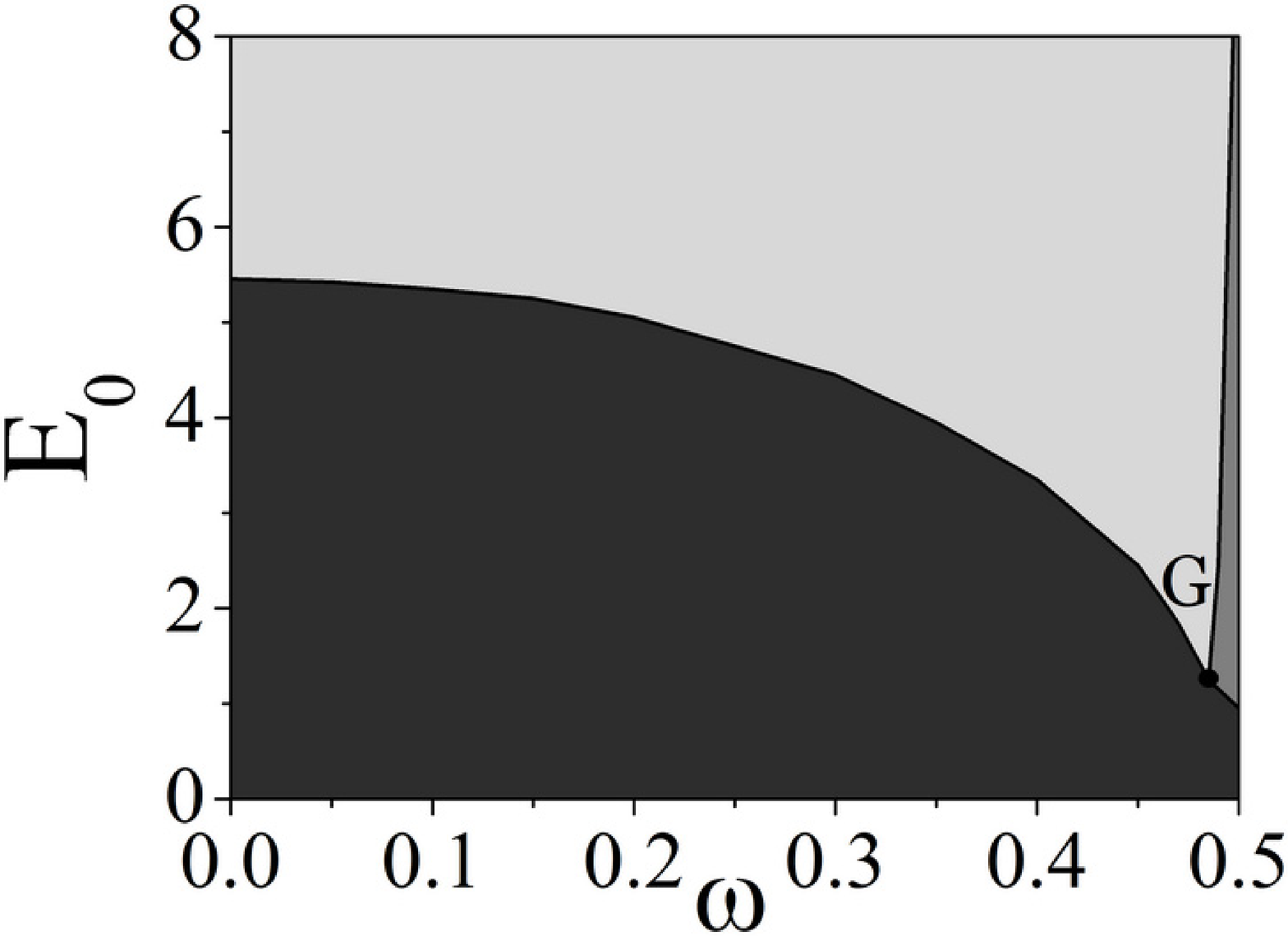}}%
\subfigure[]{ \label{fig3d}
\includegraphics[scale=0.2]{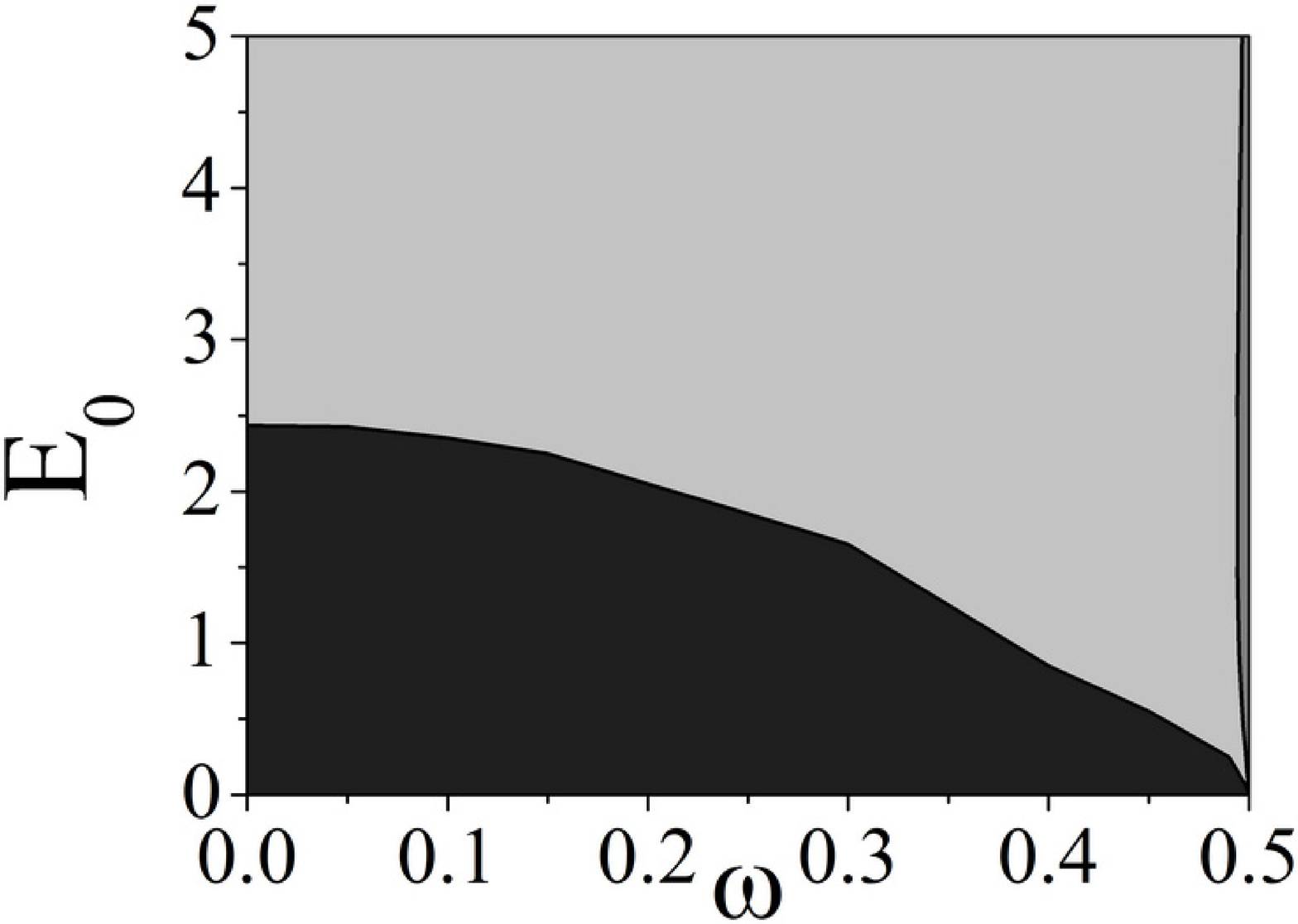}}
\caption{Stability diagrams in the plane of $(E_{0},\omega)$ inside the FRBZ. (a), (b) and (c) are for the SF case, while (d) is for the SDF case. In panel (a), (b) and (c), dark grey (bottom) represents the stability region of symmetric GS mode, light grey (top) represents the stability region of asymmetric GS mode and middle grey (right edge) represents the bistable region of these two types of modes. The parameters of them are (a) $A=2, P=0.5$, (b)$A=3,P=1$ and (c) $A=2,P=0.25$. In panel (d), dark grey (bottom) represents the stability region of antisymmetric mode, light grey (top) represents the stability region of antiasymmetric mode and middle grey (right edge) represents the bistable region of these two types of modes. The parameter of it is $A=3,P=0.5$.}
\label{Stab}
\end{figure}

Firstly, we apply the voltage $U$ along the positive direction of the extraordinary axis of BPRC, leading to $E_{0}>0$, which results in Eq. (\ref{NLS}) featuring self-focusing (SF) nonlinearity. The intensity of the induction field (alias, the pumps) is given as $I=A\sin^{2}\theta$. According to Eq. (\ref{NLS}), the maximums of the intensity induces the minimums of the linear potential at $\theta=\pm\pi/2$ (The numerical domain here is selected in $-\pi/2<\theta\leq\pi/2$), which creates the DWP along the circle. Therefore, by giving a set of parameter $(E_{0},\omega,A,P)$, we can obtained the solution from Eq. (\ref{NLS}). The mode solutions in this work are drawn from the numerical code named ``PCSOM" which is borrowed from Ref. \cite{JYang}. For the SF case, because the symmetry breaking occurs in the GS mode, other types of mode are not considered here. Typical examples of the stable symmetric and asymmetric GS mode with nonzero rotating speed are displayed in Fig. \ref{GSSF}. The figures imply that increasing $E_{0}$ leads to the symmetry breaking of the GS mode. The stability regions of these two modes drawn in the plane of $(E_{0},\omega)$ with different $A$ and $P$ in the first rotational Brillouin zone (FRBZ, the interval of $0\leq\omega\leq1/2$) \cite{YLi2} are shown in panel \ref{fig3a}-\ref{fig3c} of Fig. \ref{Stab}.


\begin{figure}[tbp]
\centering%
\subfigure[] {\label{fig4a}
\includegraphics[scale=0.2]{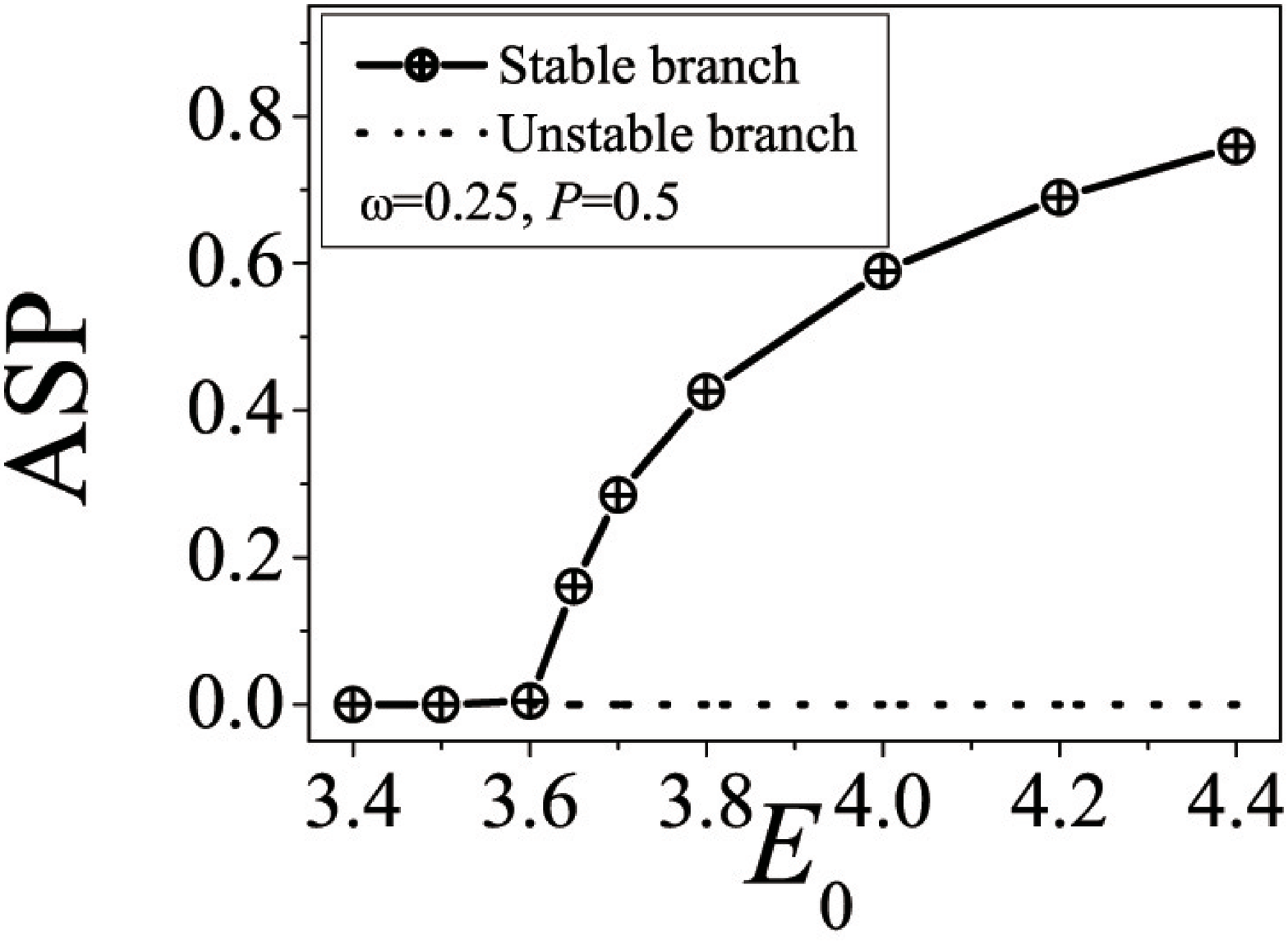}}%
\subfigure[] {\label{fig4b}
\includegraphics[scale=0.2]{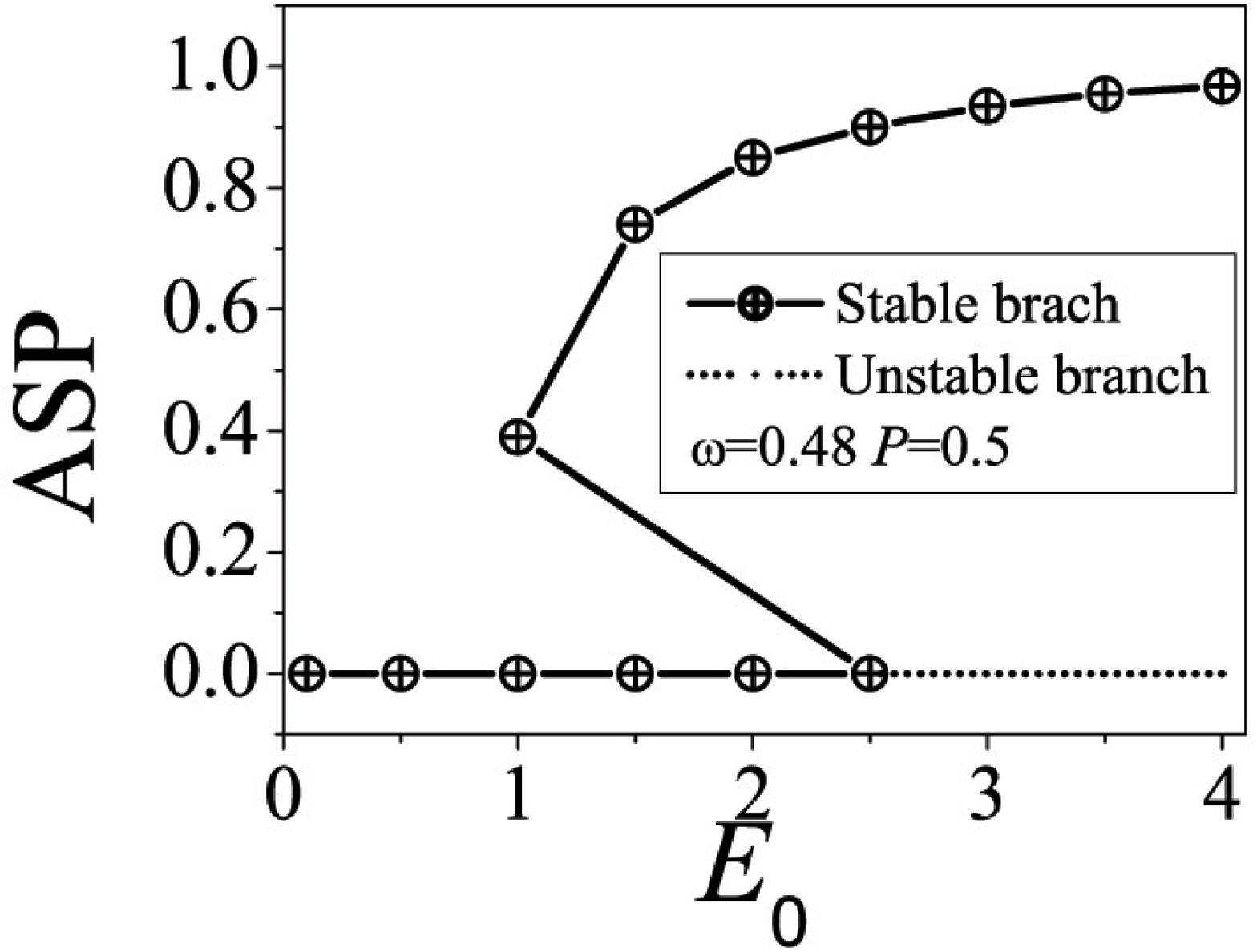}}
\caption{The ASP of the GS symmetric and asymmetric mode, defined by Eq. (\ref{ASP}), as function of the $E_{0}$ before (a) and after (b) the `G' point. The solid line represents the branch where the stable solution are found, while the dot line represent the finding of unstable solution. Panel (a) shows the bifurcation of the supercritical type, while panel (b) shows the subcritical type.} \label{Bifurcation}
\end{figure}

\begin{figure}[tbp]
\centering%
\subfigure[] {\label{fig5a}
\includegraphics[scale=0.2]{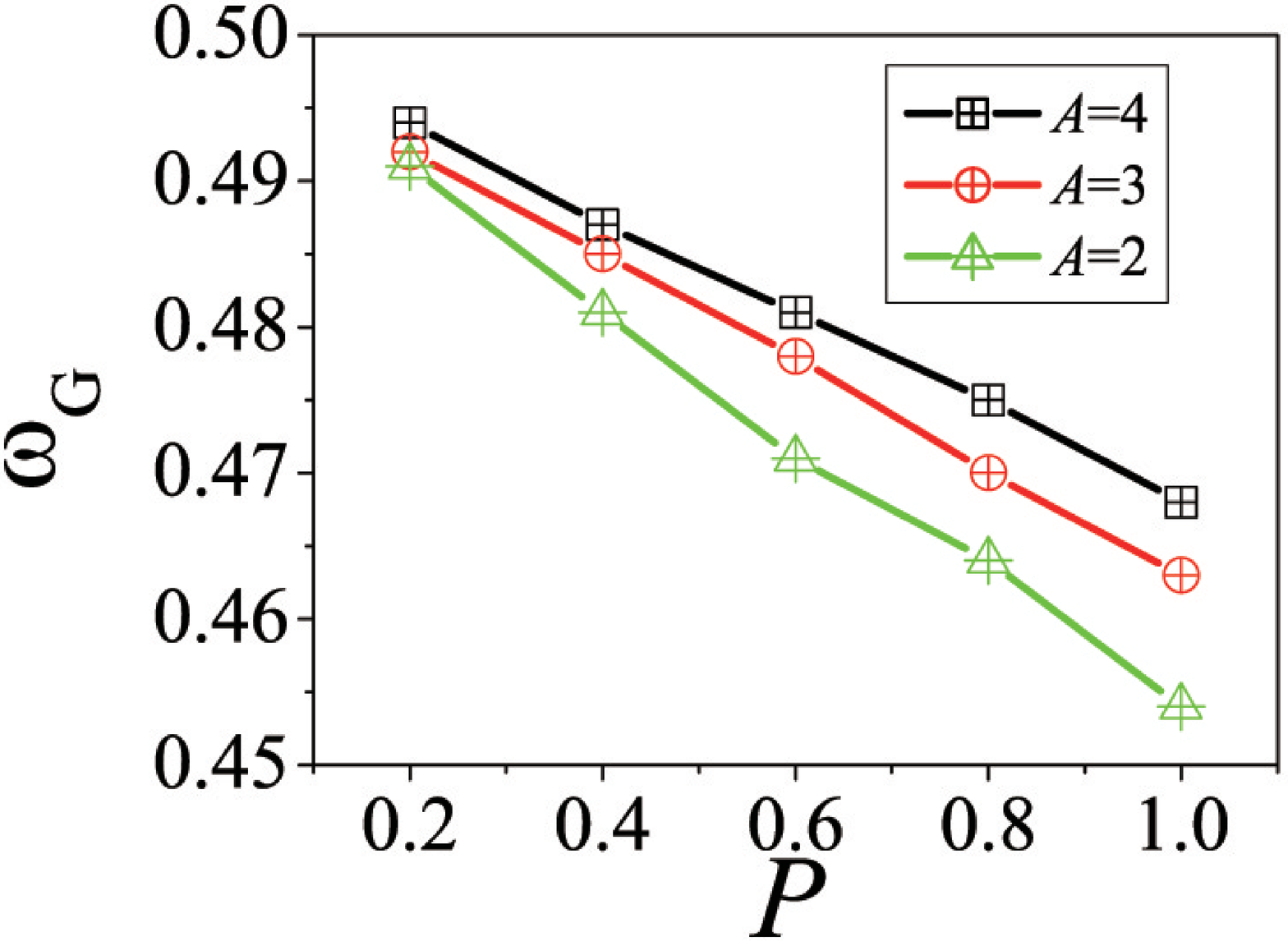}}%
\subfigure[] {\label{fig5b}
\includegraphics[scale=0.2]{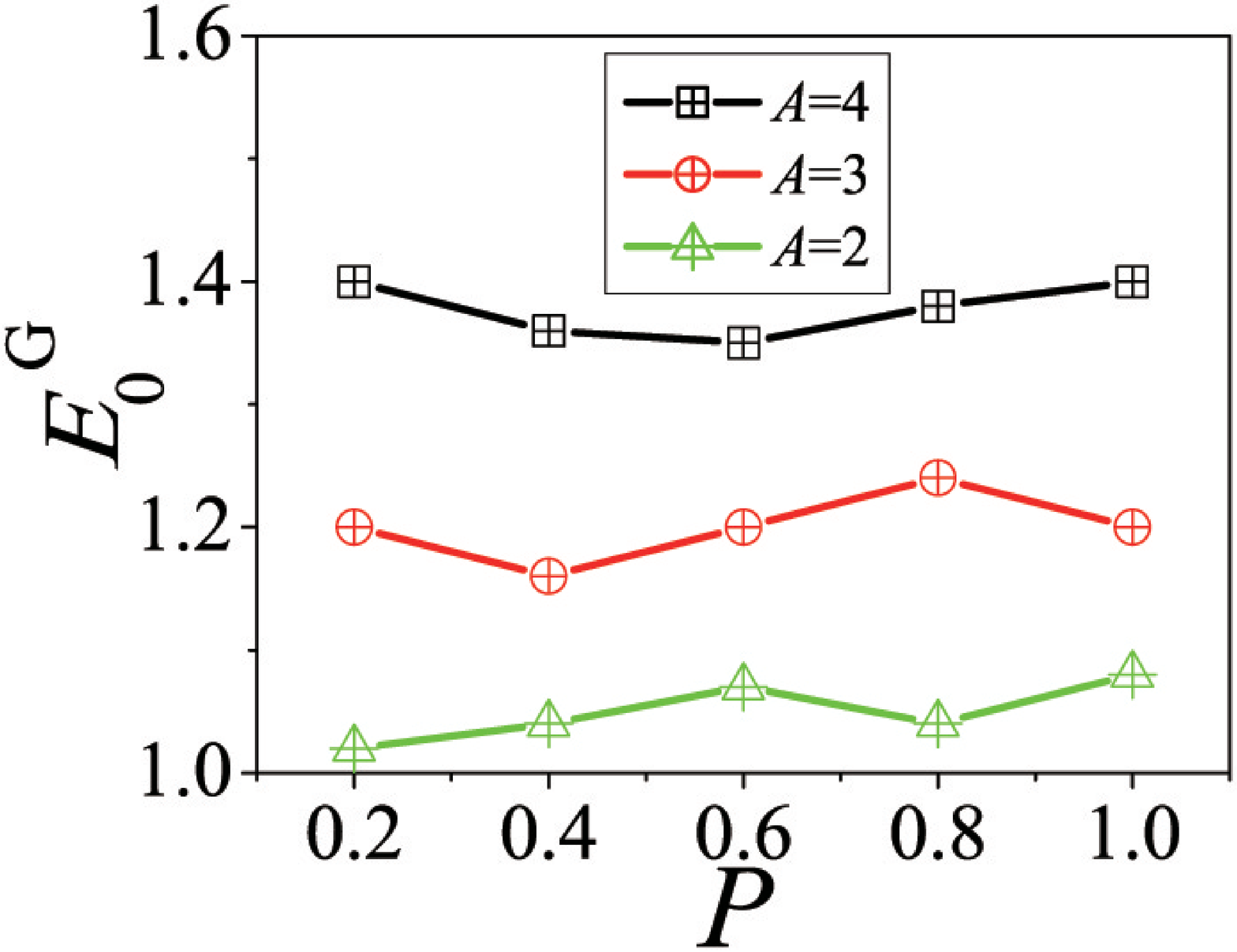}}
\caption{(Color online)(a)Coordinate $\omega_{G}$ as a function of $P$ with $A=2,3$ and $4$. (b) Coordinate $E^{G}_{0}$ as a function of $P$ with $A=2,3$ and $4$.} \label{Coordinate}
\end{figure}

In Fig. \ref{fig3a}-\ref{fig3d}, it is interesting to find a special point (labeled as `G' inside the panels) dividing the plane into three areas (symmetric area, asymmetric area and bistable area), which plays the role similar to a triple phase point. Prior to the point ($\omega<\omega_{G}$), the symmetric mode transits to the asymmetric mode directly, and afterwards ($\omega>\omega_{G}$), the transition happens after crossing the bistable area. This phenomenon implies that this point also separates the types of transition between the symmetric and asymmetric modes. To identified the symmetry breaking transition (alias SBB), we defined the asymmetry parameter (ASP) of the GS mode as
\begin{eqnarray}
\mathrm{ASP}=\left|\int^{\pi}_{0}|\phi|^{2}d\theta-\int^{0}_{-\pi}|\phi|^{2}d\theta\right|/P. \label{ASP}
\end{eqnarray}
Typical bifurcation diagrams of SBB on the either side of the `G' point expressed by the $\mathrm{ASP}(E_{0})$ are shown in Fig. \ref{Bifurcation}. The plots demonstrating the SBB prior to and after the 'G' point are the \emph{supercritical} type and \emph{subcritical} type, respectively. The `G' point, which has the rotating speed close to the right edge of the FRBZ, denotes the transition between these two kinds of SBB. The positions of point `G' in the $(E_{0},\omega)$ plane are determined by the coordinates $(E^{G}_{0},\omega_{G})$. These coordinates as a function of total power $P$ with different values of $A$ (the intensity of the pumps) are displayed in Fig. \ref{Coordinate}. The panels in Fig. \ref{Coordinate} show that larger power of the probe, $P$, leads to the transition point moving toward the left side of the plane; higher intensity of the pump leads to the point moving upward. These properties give the possibility to control the types of SBB in such rotating system.


\begin{figure}[tbp]
\centering%
\subfigure[] {\label{fig6a}
\includegraphics[scale=0.21]{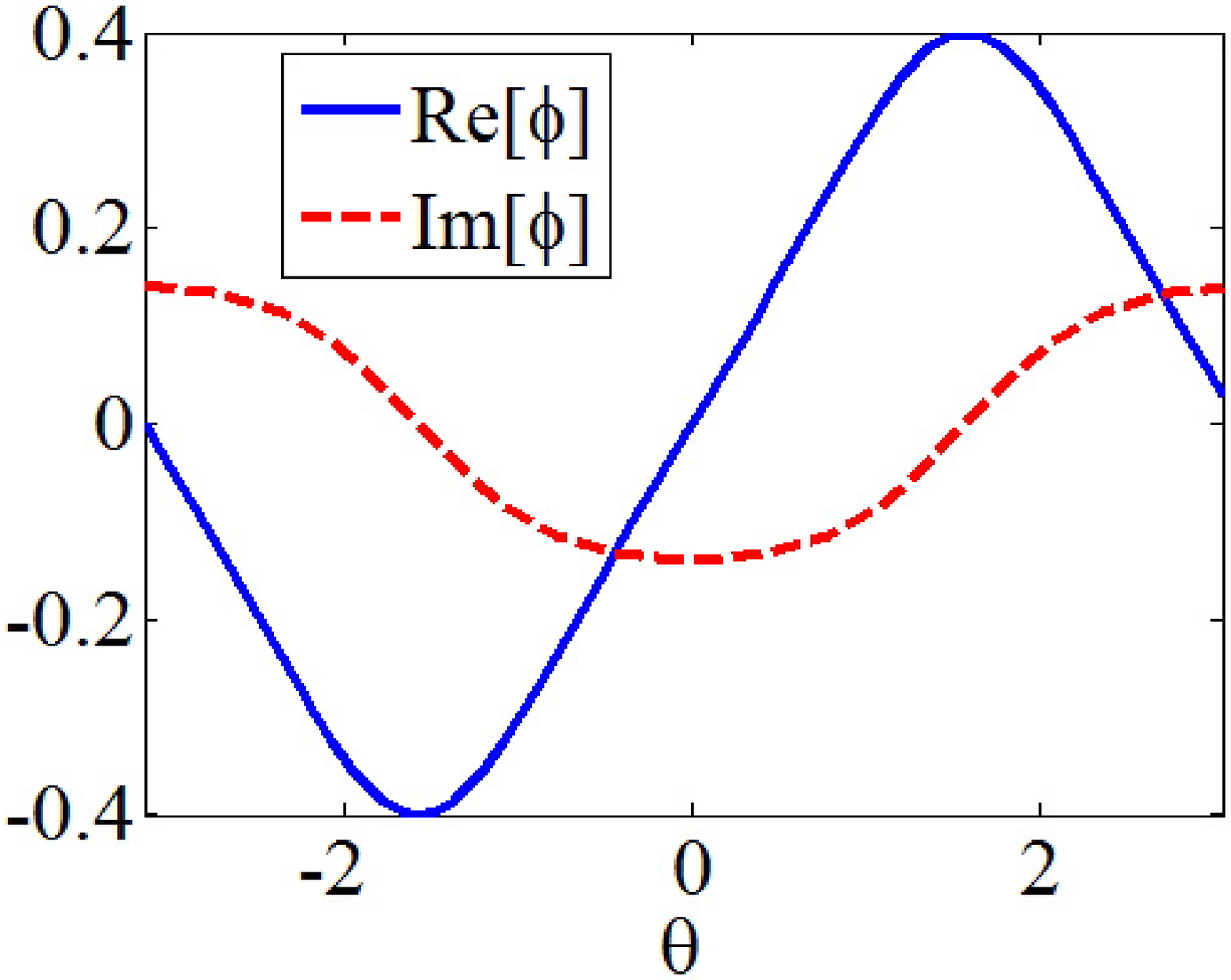}}
\subfigure[] {\label{fig6b}
\includegraphics[scale=0.21]{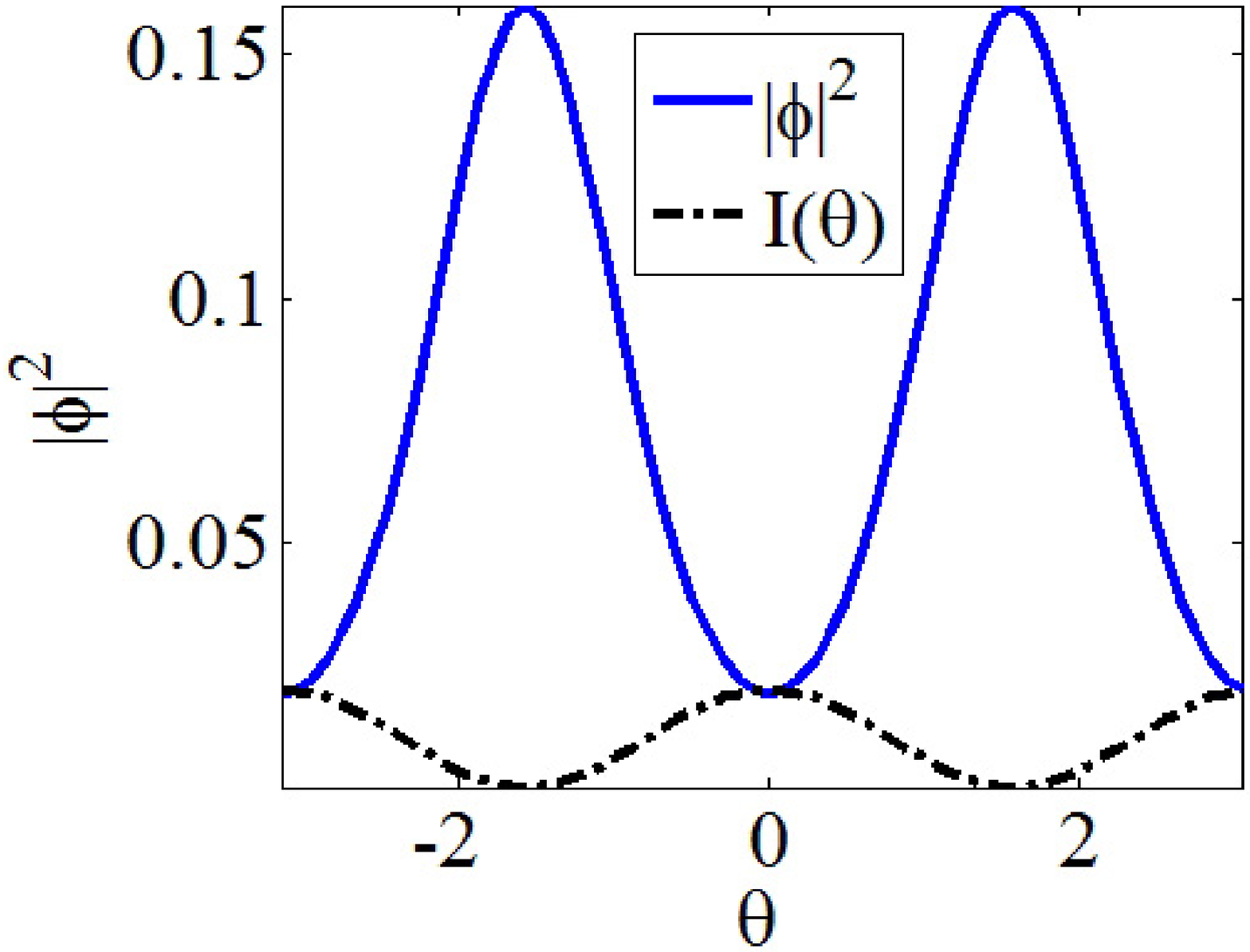}}
\subfigure[] {\label{fig6c}
\includegraphics[scale=0.21]{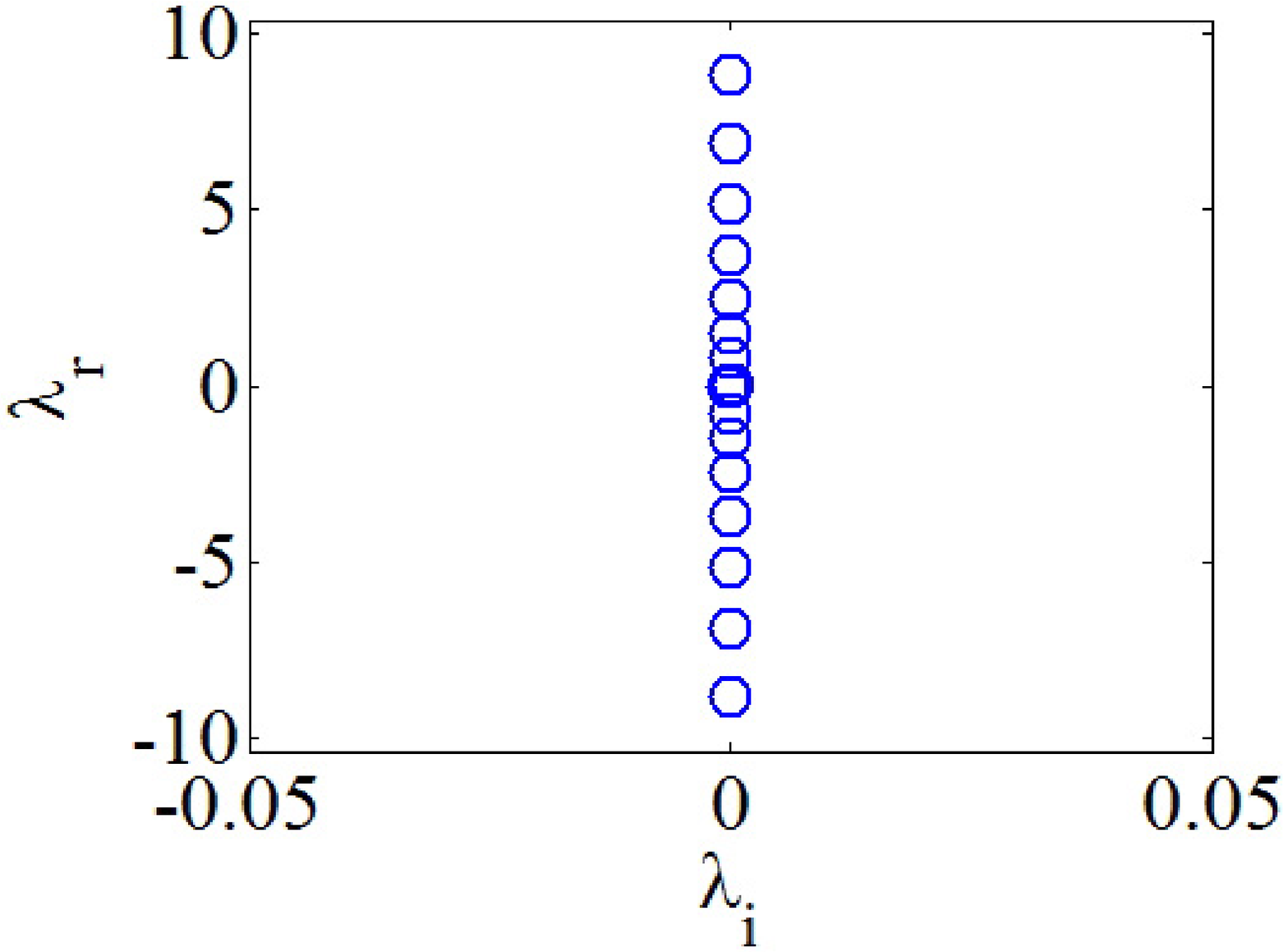}}
\subfigure[] {\label{fig6d}
\includegraphics[scale=0.21]{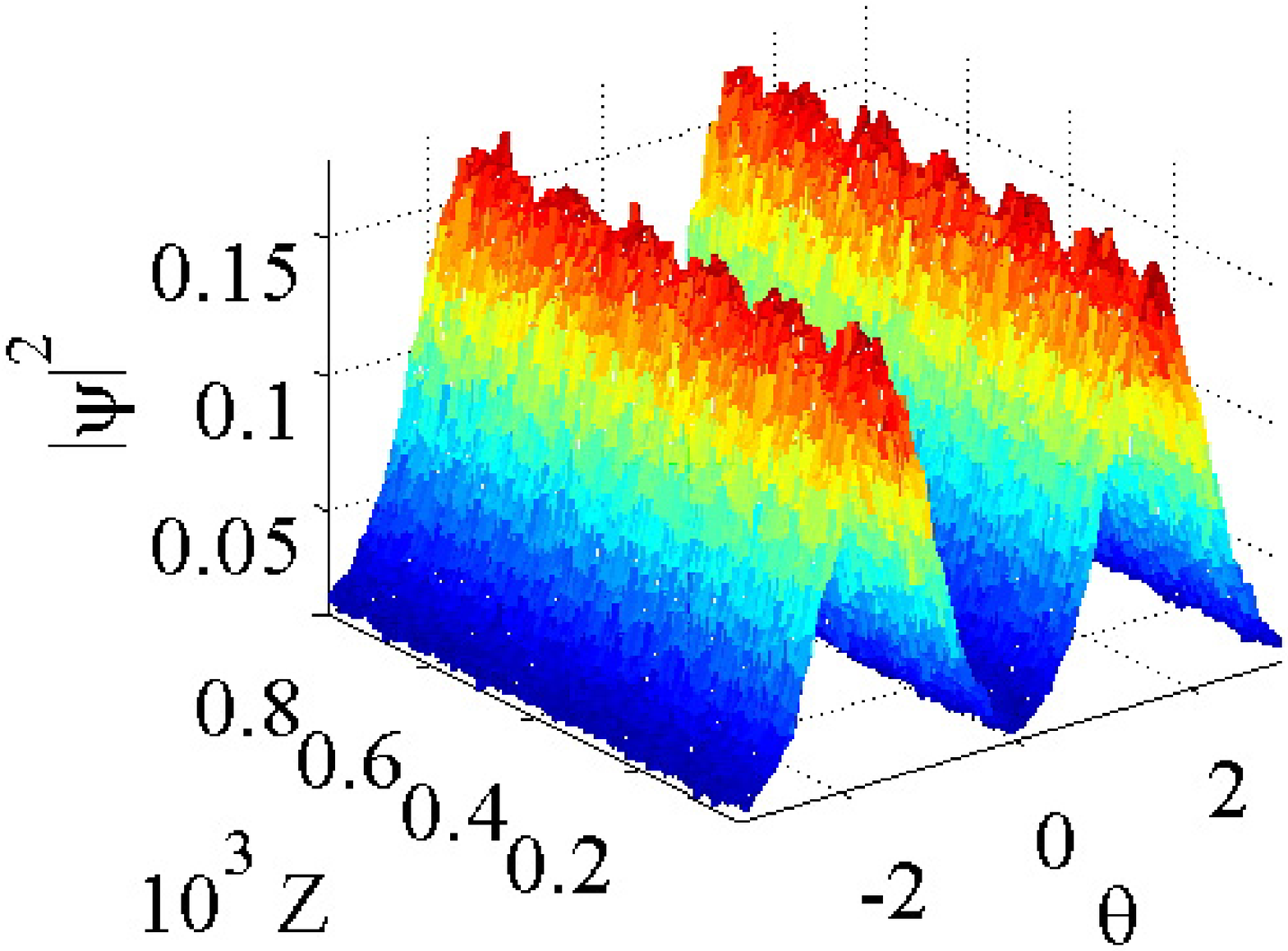}}
\subfigure[]{ \label{fig6e}
\includegraphics[scale=0.21]{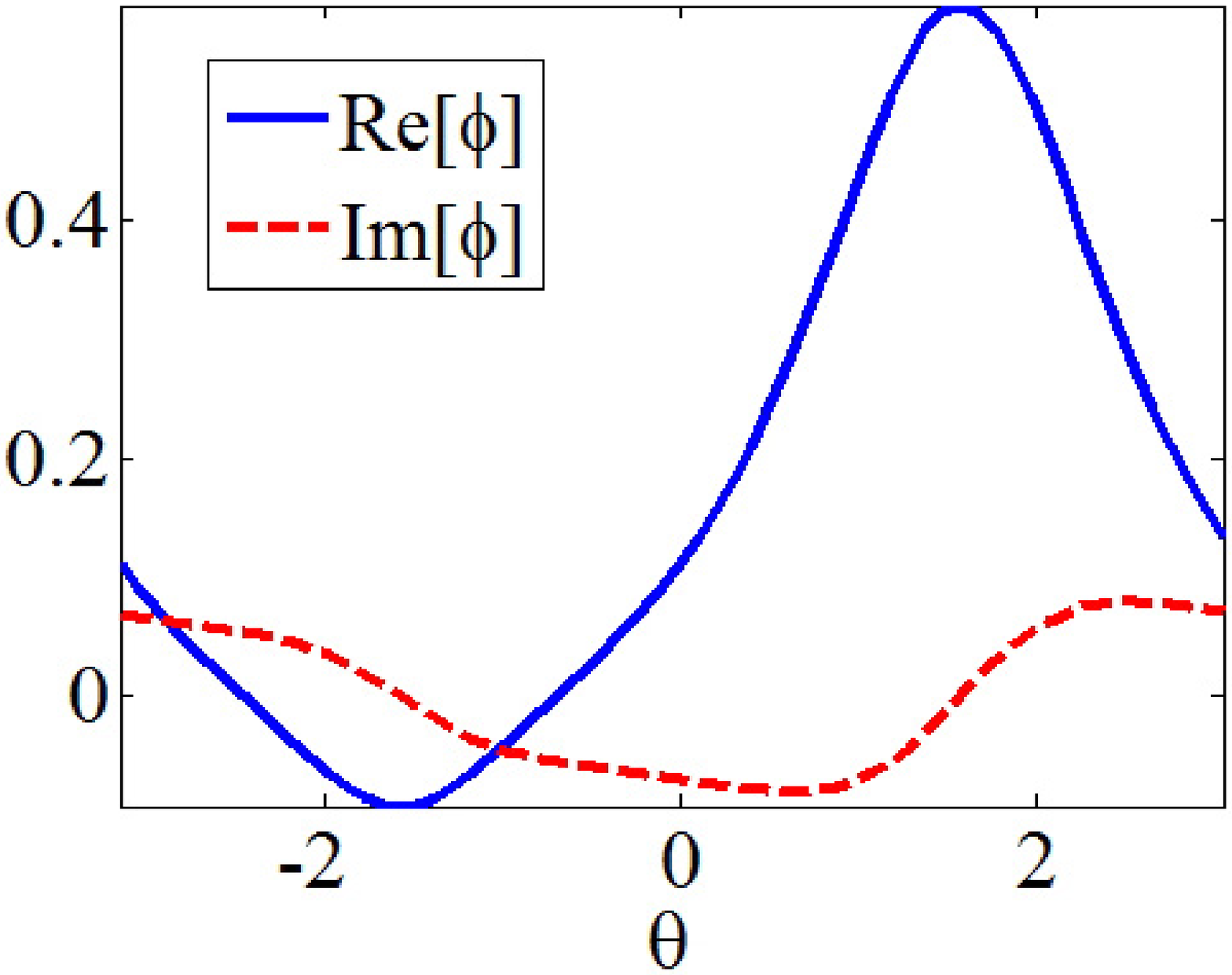}}
\subfigure[]{ \label{fig6f}
\includegraphics[scale=0.21]{2f.eps}}
\subfigure[] {\label{fig6g}
\includegraphics[scale=0.21]{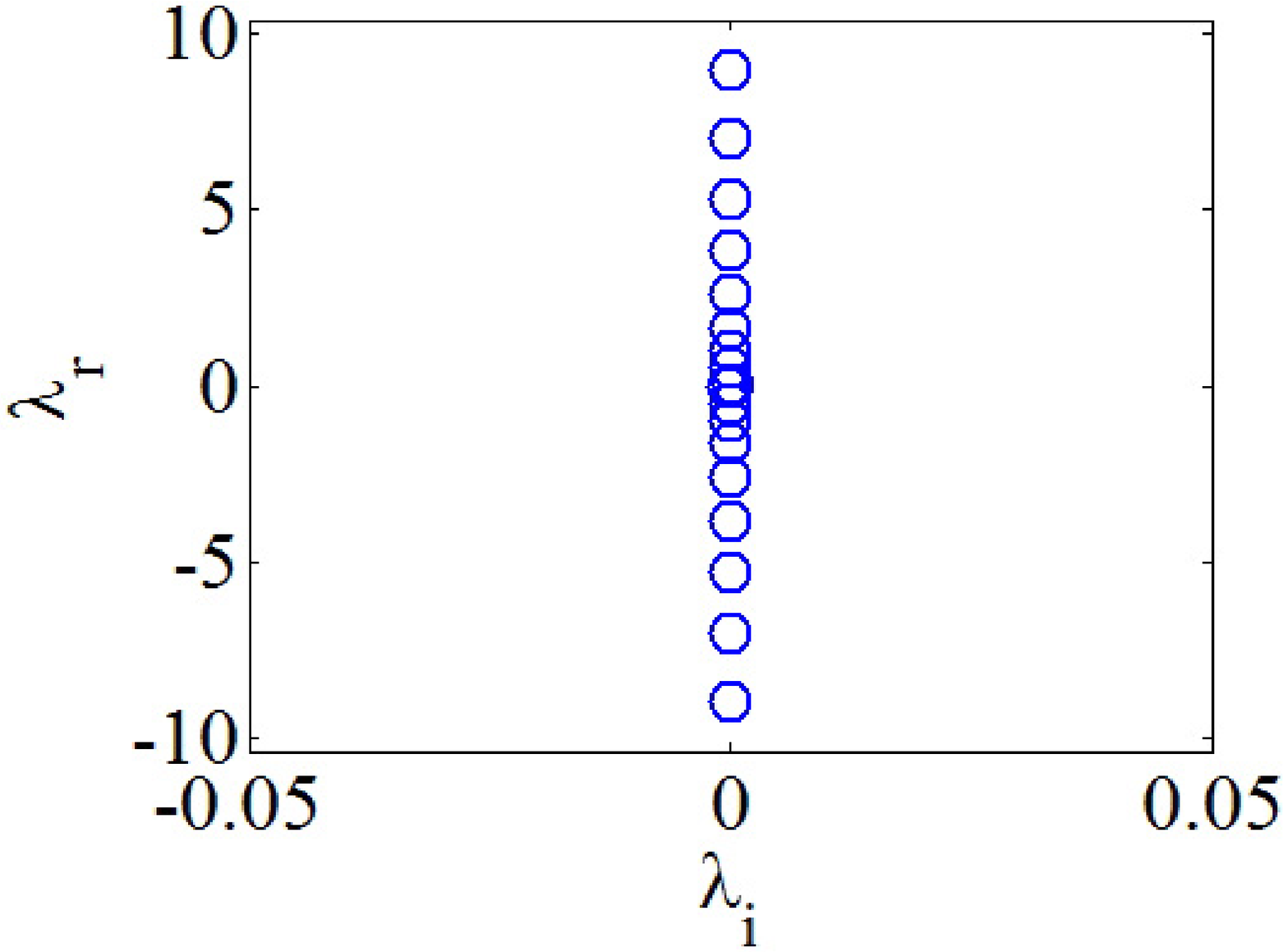}}
\subfigure[] {\label{fig6h}
\includegraphics[scale=0.21]{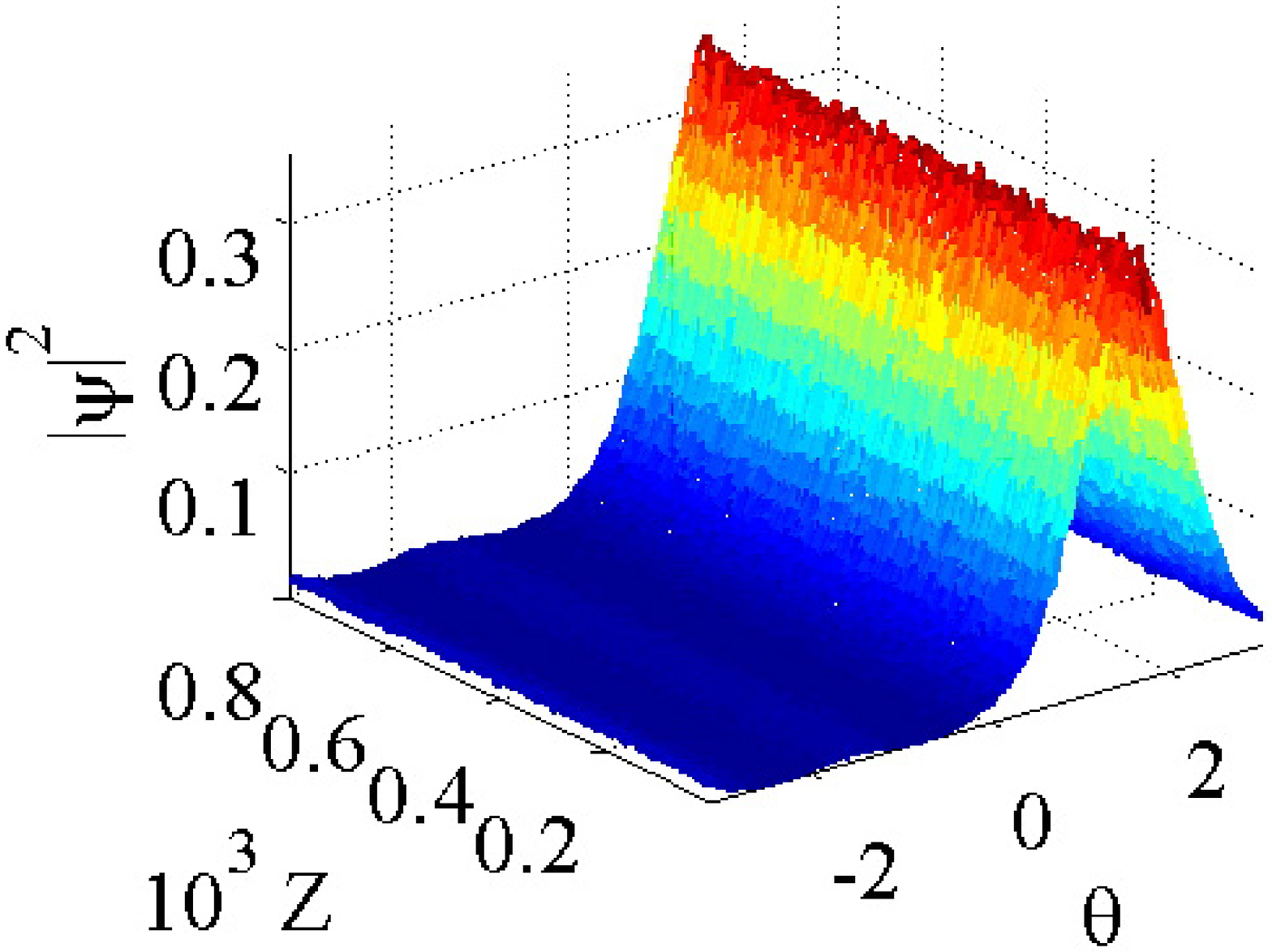}}
\caption{(Color online) Examples of stable antisymmetric and anti-asymmetric modes,
found in the system with SDF case at rotation speed $\omega =0.25$, with the parameters $(E_{0},A,P)=(2,2,0.5)$ and $(4,2,0.5)$. Panels (a), (c) display, severally, real and imaginary parts of the antisymmetric and anti-asymmetric modes, while (b), (d) show their local-power (density) profiles. The dashed-dot curves in (b) and (d) depict the corresponding intensity profile, $I(\theta)$; in the present case, it is $\cos ^{2}\protect\theta$. Panels (c), (g) exhibit the growing rates of the wave modes. And panel (d), (h) demonstrate the evolutions of these wave modes perturbed by 10\% noises, respectively. }
\label{MSDF}
\end{figure}

If we apply the voltage $U$ along the negative direction of the extraordinary axis of BPRC, it gives rise to $E_{0}<0$ and results in Eq. (\ref{NLS}) featuring self-defocusing (SDF) nonlinearity. Under this circumstance, the maximum intensity of the pump becomes the maximum of the linear potential. Therefore, in the SDF case, we select the intensity profile of the pump as $I=A\cos^{2}\theta$, which keeps the minimums of the linear potential appearing at $\theta=\pm\pi/2$ (recall that our numerical domain is select to $-\pi<\theta\leq\pi$). The symmetry breakings in SDF case are occured in the non-ground state. In the DWP, it takes place between the antisymmetric and anti-asymmetric state. Therefore, the symmetry breaking of this type is named as antisymmetry breaking. Typical examples of antisymmetric and anti-asymmetric modes are displayed in Fig. \ref{MSDF}. However, the stability plane of $(E_{0},\omega)$ in panel \ref{fig3d} shows no transition point is found. The antisymmetric mode transit to the anti-asymmetric mode directly, implying the type of SBB of antisymmetry breaking is only \emph{supercritical} \cite{Matuszewski}. Therefore, in the SDF case, we cannot switch the type of SBB.

\emph{Conclusion:} the objective of this work is to study the possibility of switching (or controlling) the type of SBB in the rotating DWP system made onto the BPRCs. The rotating potential can be created by means of optical-induction techniques, using vortex pump waves with the ordinary polarization in a photorefractive material. Increase the magnitude of the biased voltage can lead to the symmetry breaking of the mode inside the waveguide. The numerical analysis of the symmetry breaking is studied inside the FRBZ. For SF case, the symmetry breaking occurs in the GS. A special triple-like point, which separates the type of SBB (supercritical and subcritical), is found close to the right edge of the FRBZ. On the left side of the point, the type of SBB is supercritical, while on the right side of the point it is subcritical. The influence on the position of the point in the $(E_{0}, \omega)$ plane is also studied. It shows that the type of SBB can be switched in the rotating system in the SF case. For the SDF case, the type of symmetry breaking is antisymmetry breaking, which takes place between the antisymmetric and anti-asymmetric mode. The type of SBB of the antisymmetry breaking is only the supercritical type, which results in the switch of SBB cannot be achieved in this case.

This work was support by the CNNSF (grant No. 11204089,11205063). 

\bibliographystyle{plain}
\bibliography{apssamp}

\end{document}